\documentclass[10pt,leqno]{amsart}

\usepackage{amssymb,amsthm,amsmath}
\usepackage{graphicx,xcolor,paralist,tikz,fancyhdr,etoolbox,sidecap}
\usepackage[hidelinks]{hyperref}
\usepackage[hmargin=2.5cm,vmargin=2.5cm]{geometry}
\usepackage[foot]{amsaddr}

\begin{document}
\title[Modelling vehicle and pedestrian collective dynamics]{Modelling vehicle and pedestrian collective dynamics: Challenges and advances} 
\author[A. Tordeux]{Antoine Tordeux$^1$}
\address{$^1$University of Wuppertal, Germany -- \normalfont \textit{E-mail address:} \texttt{tordeux@uni-wuppertal.de}}
\author[C. Appert-Rolland]{Cécile Appert-Rolland$^2$}
\address{$^2$CNRS \& University Paris-Saclay, France}
\author[A. Nicolas]{Alexandre Nicolas$^3$}
\address{$^3$CNRS \& University Lyon 1, France}
\author[A. Seyfried]{\smallskip\\Armin Seyfried$^4$}
\address{$^4$Forschungszentrum Jülich \& University of Wuppertal, Germany}
\author[D. Ullmo]{Denis Ullmo$^5$}
\address{$^5$University Paris-Saclay, France}

\let\thefootnote\relax

\begin{abstract}
In our urbanised societies, the management and regulation of traffic and pedestrian flows is of considerable interest for public safety, economic development, and the conservation of the environment.
However, modelling and controlling the collective dynamics of vehicles and pedestrians raises several challenges. Not only are the individual entities  self-propelled and hard to describe, but their complex nonlinear physical and social interactions makes the multi-agent problem of crowd and traffic flow even more involved.
In this chapter, we purport to review the suitability and limitations of classical modelling approaches through four examples of collective behaviour:
stop-and-go waves in traffic flow,  
lane formation,
long-term avoidance behaviour, 
and load balancing in pedestrian dynamics. 
While  stop-and-go dynamics and lane formation can both be addressed by basic reactive models (at least to some extent), the latter two require anticipation and/or coordination at the level of the group. 
The results highlight the limitations of classical force-based models, but also the need for long-term anticipation mechanisms and multiscale modelling approaches.
In response, we review new developments and modelling concepts.
\end{abstract}

\maketitle

\section{Introduction}
\label{sec:Intro}

Our societies are becoming increasingly urbanised. Transport and mobility are also fundamental to the quality of everyday life. 
In this context, managing traffic flow, mixed urban traffic, and pedestrian dynamics is of considerable interest for public safety, economic development, and the conservation of the environment.
In particular,  crowd management in large buildings, such as train stations or stadiums, or during large outdoor events, such as festivals or concerts, has become the focus of much attention \cite{feliciani2022introduction}. 
However, controlling traffic flow and pedestrian dynamics is not straightforward. 
In fact, pedestrian crowding and traffic flow are collective behaviours of many-body systems subject to complex social and physical interactions. 
In practice, intelligent traffic systems and crowd management are generally designed based on simulations and computational optimisation techniques. 
Simulations are based on pedestrian and traffic flow modelling approaches, and the demand for reliable and realistic models is getting more and more pressing. 
However, much of the collective drivers' and pedestrians' behaviour remains poorly understood. 
For example, traffic flow stabilisation is still a challenge, especially in the context of driving automation \cite{gunter2020commercially}, while the notion of anticipation for a pedestrian is still somewhat fuzzy.
 
In congested conditions, pedestrian and road traffic flows essentially describe collective dynamics and coordinated behaviours. 
These coordinated dynamics can result from a form of collective intelligence, for example, when pedestrian counterflows self-organise into lanes to optimise flow \cite{helbing2005self} (although this phenomenon is also observed in passive particle systems such as colloids with different directions of motion),  when crowds clear the way to facilitate the passage of an \emph{intruder} \cite{nicolas2019mechanical,bonnemain2023pedestrians}, for example an ambulance (e.g., in the video \cite{MadrasAmbulance}), or when pedestrians spread out over different possible exits to balance the load on each exit during evacuation \cite{wagoum2017understanding}. 
That being said, collective dynamics are not always beneficial and can even be strongly detrimental. Consider the stop-and-go waves (accordion traffic) commonly observed on congested motorways, which negatively impact traffic safety and performance and heighten the environmental harm caused by traffic because of the increased gas consumption due to repeated accelerations and braking.
For all these problems, modelling, understanding and controlling collective dynamics starting from the local interactions of the agents from a complex systems perspective is a challenging task \cite{bellomo2012modeling}.

\subsection{Modelling vehicle and pedestrian collective dynamics}

The modelling of vehicle and pedestrian collective dynamics is generally divided into three different levels: strategic, tactical, and operational. 
This concept, that originally finds its roots in the military field, was introduced to the modelling of driver behaviour in the 1970s by Michon \cite{michon1985critical} and was transferred to pedestrian dynamics in the early 2000s \cite{hoogendoorn2004pedestrian}. 
At the strategic level, the modeller is concerned with the largest scales, those of activity choices and scheduling (departure time planning). 
The focus of the tactical level is on smaller scales; it includes manoeuvre planning, the choice of a path, or the choice of an exit in the context of an evacuation. Operational modelling describes local motion in interaction with neighbours and obstacles. 
The state of the art generally treats the strategic, tactical and operational modelling scales separately. 
However, we argue in this review that these classifications are not entirely distinct from one another, but could be strongly interlaced, especially the operational and tactical modelling scales. 
This leads to conceptual problems in modelling these systems. 
Among operational models, three categories are classically distinguished: microscopic, mesoscopic, and macroscopic approaches \cite{chowdhury2000statistical,helbing2001traffic}. 
Microscopic models deal with the individual motion of the agents in Lagrangian coordinates, while mesoscopic and macroscopic approaches aggregate the dynamics in Eulerian coordinates using probabilistic agent distribution functions and averaged quantities such as flow and density, respectively. 
Among these approaches, first-order models posit that the agent's speed \emph{instantaneously} adapts to the environment, 
typically using proactive models based on anticipation and optimisation, whereas force-based models are second-order approaches that are formally similar to Newton's equation and thus incorporate into the dynamics `inertia', i.e., gradual adjustments of an agent's velocity to the current situation, along with possible mechanical forces.

Microscopic operational models reproduce the outcome of competition between conflicting goals. Agents are driven towards a destination defined at the tactical and strategic modelling levels, but at the same time they are intent on avoiding collisions with other entities and unfavourable regions in space. The resulting interactions between entities need not obey Newton's third law of motion, where action equals reaction, because vehicles and pedestrians self-propel in a direction set by their own decisions, rather than the resultant of mechanical forces. In addition, interactions are generally strongly anisotropic, the presence of neighbours or obstacles being much more dominant when located in the direction of motion, and nonlinear, through short-range interaction models or finite dynamic capacities.
The analysis of such systems can be extremely tedious, particularly when considering nonlinear anisotropic interaction models. 
In addition, many situations require explicit anticipation on the long-term and on-the-fly decisions about collective tactics, resulting in dynamic choices at different modelling scales. 

\subsection{Objectives and organisation of the chapter}

In this chapter, we purport to review the suitability of vehicle and pedestrian models to describe phenomena that seem to pertain to coordinated collective dynamics.
Specifically, we highlight that current models fail to consider goals beyond the operational movement of a single pedestrian and represent a collective goal of a group or subgroup. 
These are typical limitations of classical reactive approaches with pairwise interaction, motivating the need for collective prediction and proactive models that exhibit long-term anticipation and coordination with complex interaction schemes coupling operational and tactical scales, among other modelling concepts. 
We study four examples of collective phenomena: 1) stop-and-go waves in traffic flow, 2) lane formation for pedestrian counterflow, 3) long-term anticipation in crowd crossed by an intruder, and 4) load balancing between multiple exits in the context of an evacuation. 
While both stop-and-go and lane formation collective phenomena can be addressed by basic reactive models, the latter require proactive control, optimisation, and collective coordination. 
The chapter is organised as follows. 
Section~\ref{sec:Models} provides a concise overview of the state-of-the-art in operational vehicle and pedestrian models in one and two spatial dimensions.
The focus is on continuous approaches using ordinary and delay differential equation systems.
In section~\ref{sec:CollDyn}, we review the four collective behaviours of stop-and-go waves, lane formation, coordinated obstacle avoidance, and load balancing in light of current modelling approaches and concepts. 
Finally, in section~\ref{sec:Challenges}, we discuss some challenges and limitations of current models and propose some development perspectives and research directions.

\section{Vehicle and pedestrian models}
\label{sec:Models}

We start by briefly reviewing continuous vehicle and pedestrian operational models using differential systems. 
The review includes car-following in one dimension in section~\ref{subsec:CF-Models} and pedestrian models in two dimensions in section~\ref{subsec:Ped-Models}, covering force-based reactive models and anticipation-based proactive approaches.
We refer to the books \cite{cristiani2014multiscale,maury2018crowds,schadschneider2010stochastic,treiber2013traffic} for more in-depth reviews of the topic, including mesoscopic and macroscopic modelling approaches.

\subsection{Car-following models}
\label{subsec:CF-Models}

\begin{figure}[!ht]
    \centering
\begin{tikzpicture}[x=1.2pt,y=1.2pt]
\definecolor{fillColor}{RGB}{255,255,255}
\path[use as bounding box,fill=fillColor,fill opacity=0.00] (10,0) rectangle (227.62, 71.13);
\begin{scope}
\path[clip] (  10.00,  0.00) rectangle (227.62, 71.13);
\definecolor{drawColor}{gray}{0.20}

\path[draw=drawColor,line width= 0.8pt,line join=round,line cap=round] ( 28.19, 16.75) -- ( 67.71, 16.75);

\path[draw=drawColor,line width= 0.8pt,line join=round,line cap=round] ( 28.19, 16.75) -- ( 28.19, 26.16);

\path[draw=drawColor,line width= 0.8pt,line join=round,line cap=round] ( 67.71, 16.75) -- ( 67.71, 26.16);

\path[draw=drawColor,line width= 2.0pt,line join=round,line cap=round] ( 67.71, 21.45) -- ( 67.71, 24.28);

\path[draw=drawColor,line width= 0.8pt,line join=round,line cap=round] ( 28.19, 26.16) -- ( 67.71, 26.16);

\path[draw=drawColor,line width= 0.8pt,line join=round,line cap=round] ( 31.48, 26.16) -- ( 41.36, 35.57);

\path[draw=drawColor,line width= 0.8pt,line join=round,line cap=round] ( 41.36, 35.57) -- ( 54.53, 35.57);

\path[draw=drawColor,line width= 0.8pt,line join=round,line cap=round] ( 54.53, 35.57) -- ( 61.12, 26.16);

\path[draw=drawColor,line width= 0.8pt,line join=round,line cap=round] ( 61.12, 26.16) -- ( 67.71, 26.16);
\definecolor{fillColor}{gray}{0.20}

\path[fill=fillColor] ( 38.07, 16.75) circle (  4.50);

\path[fill=fillColor] ( 57.83, 16.75) circle (  4.50);

\path[draw=drawColor,line width= 0.8pt,line join=round,line cap=round] (159.92, 16.75) -- (199.43, 16.75);

\path[draw=drawColor,line width= 0.8pt,line join=round,line cap=round] (159.92, 16.75) -- (159.92, 26.16);

\path[draw=drawColor,line width= 0.8pt,line join=round,line cap=round] (199.43, 16.75) -- (199.43, 26.16);

\path[draw=drawColor,line width= 2.0pt,line join=round,line cap=round] (199.43, 21.45) -- (199.43, 24.28);

\path[draw=drawColor,line width= 0.8pt,line join=round,line cap=round] (159.92, 26.16) -- (199.43, 26.16);

\path[draw=drawColor,line width= 0.8pt,line join=round,line cap=round] (163.21, 26.16) -- (173.09, 35.57);

\path[draw=drawColor,line width= 0.8pt,line join=round,line cap=round] (173.09, 35.57) -- (186.26, 35.57);

\path[draw=drawColor,line width= 0.8pt,line join=round,line cap=round] (186.26, 35.57) -- (192.85, 26.16);

\path[draw=drawColor,line width= 0.8pt,line join=round,line cap=round] (192.85, 26.16) -- (199.43, 26.16);

\path[fill=fillColor] (169.79, 16.75) circle (  4.50);

\path[fill=fillColor] (189.55, 16.75) circle (  4.50);

\path[draw=drawColor,line width= 0.4pt,line join=round,line cap=round] ( 47.95, 12.04) -- ( 47.95,  9.22);

\path[draw=drawColor,line width= 0.4pt,dash pattern=on 1pt off 3pt ,line join=round,line cap=round] ( 47.95, 12.04) -- ( 47.95, 56.27);

\path[draw=drawColor,line width= 0.4pt,line join=round,line cap=round] (179.67, 12.04) -- (179.67,  9.22);

\path[draw=drawColor,line width= 0.4pt,dash pattern=on 1pt off 3pt ,line join=round,line cap=round] (179.67, 12.04) -- (179.67, 56.27);

\path[draw=drawColor,line width= 0.8pt,line join=round,line cap=round] ( 47.95, 56.27) -- (179.67, 56.27);

\path[draw=drawColor,line width= 0.8pt,line join=round,line cap=round] ( 47.95, 59.16) --
	( 47.95, 56.27) --
	( 47.95, 53.37);

\path[draw=drawColor,line width= 0.8pt,line join=round,line cap=round] (179.67, 53.37) --
	(179.67, 56.27) --
	(179.67, 59.16);









\path[draw=drawColor,line width= 0.8pt,line join=round,line cap=round] ( 47.95, 40.27) -- ( 79.12, 40.27);

\path[draw=drawColor,line width= 0.8pt,line join=round,line cap=round] ( 76.12, 38.54) --
	( 79.12, 40.27) --
	( 76.12, 42.00);
\definecolor{drawColor}{RGB}{0,0,0}

\node[text=drawColor,anchor=base,inner sep=0pt, outer sep=0pt, scale=  0.90] at ( 54.53, 45.80) {~$\dot x_n$};
\definecolor{drawColor}{gray}{0.20}

\path[draw=drawColor,line width= 0.8pt,line join=round,line cap=round] (179.67, 40.27) -- (210.85, 40.27);

\path[draw=drawColor,line width= 0.8pt,line join=round,line cap=round] (207.84, 38.54) --
	(210.85, 40.27) --
	(207.84, 42.00);
\definecolor{drawColor}{RGB}{0,0,0}

\node[text=drawColor,anchor=base,inner sep=0pt, outer sep=0pt, scale=  0.90] at (190, 45.80) {~~~~$\dot x_{n+1}$};

\node[text=drawColor,anchor=base,inner sep=0pt, outer sep=0pt, scale=  0.90] at ( 47.95,  2.68) {$x_n$};

\node[text=drawColor,anchor=base,inner sep=0pt, outer sep=0pt, scale=  0.90] at (179.67,  3.39) {$x_{n+1}$};
\definecolor{drawColor}{gray}{0.20}

\path[draw=drawColor,line width= 0.8pt,line join=round,line cap=round] (  8.43, 12.04) -- (219.19, 12.04);

\path[draw=drawColor,line width= 0.8pt,line join=round,line cap=round] (212.93,  8.43) --
	(219.19, 12.04) --
	(212.93, 15.66);
\definecolor{drawColor}{RGB}{0,0,0}

\node[text=drawColor,anchor=base,inner sep=0pt, outer sep=0pt, scale=  0.90] at (113.81, 48) {$\Delta x_n=x_{n+1}-x_n$};

\end{scope}
\end{tikzpicture}
    \caption{Main variables for a car-following situation in one dimension. The dynamics of the vehicle $n$ are totally asymmetric. They solely depend on the distance and speed of the predecessor $n+1$.}
    \label{fig:scheme_CF}
\end{figure}
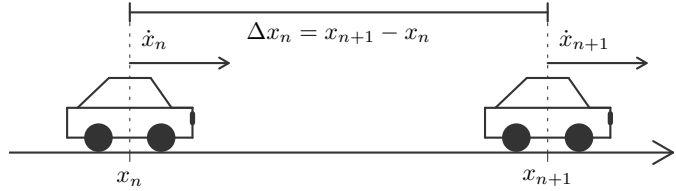

Car-following models are totally asymmetric interaction models, where the dynamics of a vehicle $n$ depends only on the nearest vehicle ahead $n+1$ (see figure~\ref{fig:scheme_CF}). 
With $x_n(t)\in\mathbb R$ the position of the vehicle $n$ at time $t$, $\Delta x_n(t)=x_{n+1}(t)-x_n(t)$ the distance to the predecessor and $\Delta \dot x_n(t)=\dot x_{n+1}(t)-\dot x_n(t)$ the speed difference, ordinary car-following models are of the form
\begin{align}
    &\dot x_n(t) = V(\Delta x_n(t),\Delta \dot x_n(t)), \label{eq:CF-speed}\\[1mm]
    &\ddot x_n(t) = A(\Delta x_n(t),\Delta \dot x_n(t),\dot x_n(t)). \label{eq:CF-acc}
\end{align}
Here \eqref{eq:CF-speed} is a first order speed model while \eqref{eq:CF-acc} is a second order acceleration model. 
Both models admit for any distance $d$ at least one equilibrium speed $v(d)$ such that $V(d,0)=v(d)$ and $A(d,0,v(d))=0$.

Car-following models have gone through two major periods of development. 
The first dates back to their inception in the 1950s and the craze that followed. 
The second, more recent, dates back to the 1990s and is still in use today.

\subsubsection{Pioneering follow-the-leader models (1950-1970)}

The first studies on car-following behaviour can be traced back to the 1950s and the pioneering work by Reuschel on the dynamics of road vehicles \cite{reuschel1950fahrzeugbewegungen}. 
In this first car-following model, the forerunner of a long series of works, the speed is directly proportional to distance to the vehicle in front
\begin{equation}
    \dot x_n(t)=\frac1T\big(\Delta x_n(t)-\ell\big),\qquad T>0,~~\ell\ge0,
\end{equation}
where $T$ is the time gap and  $\ell$ the vehicle size.
For this first-order linear model, all the states of the form $(d,v(d)=(d-\ell)/T)$ are systematically stable equilibrium states. 
The model proposed  by Pipes a few years later is also linear, but of the second order
\begin{equation}
    \ddot x_n(t)=\lambda\Delta \dot x_n(t),\qquad \lambda>0,
\end{equation}
where $\lambda$ is here a speed relaxation time. 
In this model, there is no regulation of distance as a function of velocity, and any uniform speed configuration is systematically stable. 
The end of the 1950s saw the fruitful development of delayed linear models and collective stability analysis \cite{chandler1958traffic,herman1959traffic,kometani1958stability}.
Nonlinear models date back to early 1960s with, e.g., the general \emph{Stimulus-Response} follow-the-leader model by Gazis et al.\ \cite{gazis1961nonlinear}
\begin{equation}
    \ddot x_n(t+\tau)=\lambda\dot x_{n+1}^{\,\alpha}(t+\tau)\frac{\Delta \dot x_n(t)}{\Delta x_n^{\,\beta}(t)},\qquad \tau,\lambda,\alpha,\beta>0,
\end{equation}
where $\alpha$ and $\beta$ are some exponent parameters, and the first-order exponential delayed model of Newell \cite{newell1961nonlinear}
\begin{equation}
    \dot x_n(t+\tau)=v_0\left(1-\exp\left(-\frac{\Delta x_n(t)-\ell}{Tv_0}\right)\right),\quad v_0,T>0, ~~\ell\ge0.
\end{equation}

This first series of models highlighted three main characteristic aspects of car-following behaviour:
\begin{enumerate}
    \item the need for nonlinearity in models so as to account for the two regimes of (free and congested) traffic 
    \item the effect of reaction and relaxation times on traffic stability -- 
it became clear that the models must be sufficiently reactive to achieve traffic stability
    \item the possibility to identify parameters such as the maximum speed (or desired speed) $v_0$ for free traffic, and the reaction time $\tau$ and the time gap $T$ that prevail in congested traffic. 
\end{enumerate}
In fact, in congested traffic, the speed of the vehicles is assumed to be proportional to the spacing ahead, the time gap $T$ being the inverse of the proportionality factor.
It corresponds to the time before collision between a vehicle and its predecessor if the latter were to stop abruptly while the speed of the follower remains constant.
The \emph{constant-time-gap-policy} is a following strategy that was initiated in the \emph{California Code} in the US \cite{chandler1958traffic} and physically related to safety distances in the early 1980s \cite{gipps1981behavioural}.
The policy is now widely accepted for modelling driver behaviour and also for Adaptive Cruise Control (ACC) systems (see the pioneering work \cite{swaroop1994comparision}) that is recommended in ISO standards (such as the ISO15622 with a time gap gap ranging from 0.8 to 2.2~s \cite{iso2018intelligent} or again 1.8~s for the German ADAC guidelines).

\subsubsection{Modern car-following models (1990-today)}

A second spate of nonlinear models emerged in the 1990s.
In parallel, the terminology \emph{follow-the-leader} was on the wane, in favour of \emph{car-following} (see \cite[Fig.~2]{cordes2023single}). 
This second spate is often referred to as \emph{Optimal Velocity} (OV) models. 
The concept of \emph{Optimal Velocity} comes from the systematic use of a nonlinear equilibrium speed function depending on the distance to the preceding vehicle, a notion initially introduced in the pioneering models by Reuschel \cite{reuschel1950fahrzeugbewegungen}, Chandler \cite{chandler1958traffic} and Newell \cite{newell1961nonlinear}. 
Note that the terminology \emph{Optimal Velocity} can be misleading, as there is no optimisation process here. That said, it has made its way into the literature.
In modern OV models, the OV function $V:\mathbb R\mapsto\mathbb R_+$ can take the form of a sigmoidal arc-tangent as in the seminal \emph{Optimal Velocity} model of Bando et al.\ \cite{bando1995dynamical}
\begin{equation}
    \ddot x_n(t)=\frac1\tau\big[V\big(\Delta x_n(t)\big)-\dot x_n(t)\big],\qquad \tau>0,~~V\in C^1(\mathbb R,\mathbb R_+),
    \label{eq:OVM}
\end{equation}
or the \emph{Full Velocity Difference} (FVD) model by Jiang et al.\ \cite{jiang2001full}
\begin{equation}
    \ddot x_n(t)=\frac1{\tau_1}\big[V\big(\Delta x_n(t)\big)-\dot x_n(t)\big]+\frac{\Delta \dot x_n(t)}{\tau_2},\quad \tau_1,\tau_2>0,~~V\in C^1(\mathbb R,\mathbb R_+)
    \label{eq:FVDM}
\end{equation}
including two relaxation rates: $\tau_1$ is associated with the relaxation to the distance-dependent OV, whereas the speed difference with the predecessor relaxes over a characteristic time $\tau_2$. 
This double dependence is not surprising; it is sensible to assume that car-following behaviour  depends not only on the distance to the preceding car, but also on their speed difference. 
Nowadays all modern car-following models depend on both of these  variables and also on individual speeds, as \emph{inter alia} in the famous \emph{Intelligent Driver} (ID) model by Treiber et al.\  \cite{treiber2000congested}
\begin{equation}
    \left\{~\begin{array}{lcl}
\displaystyle\ddot x_n(t)=a\left[1-\Big(\frac{\dot x_n(t)}{v_0}\Big)^4-\left(\frac{f(\dot x_{n+1}(t),\dot x_n(t))}{\Delta x_n(t)-\ell}\right)^2\right],\\[4mm]
f(v,v_1)=\ell+Tv+v\frac{v-v_1}{2\sqrt{ab}}+\Delta\sqrt{v/v_0},\end{array}\right.
\end{equation}
where $a,b,v_0,T>0$ and $\ell\ge0$ are the model parameters, with $a$ a maximum acceleration and $b$ a comfortable deceleration. 

The emergence of new nonlinear models in the 1990s and early 2000s notably followed from simulation results showing realistic stop-and-go waves in stationary states for fine-tuned parameter settings \cite{bando1995dynamical,bando1998analysis,jiang2001full}. 
The mathematical analysis of the models yielded collective stability conditions, typically $2\tau< 1/V'(d)=:T$ for the OV model \cite{bando1995dynamical}. 
In other words, the reaction time $\tau$ must be sufficiently small compared to the time gap $T$, that is, the drivers must be sufficiently reactive. 
Such findings opened the door to the modelling and control of traffic jam and stop-and-go dynamics \cite{wilson2011car,orosz2010traffic}. 
A more recent revival comes from the advent of driving automation and the need to develop stable models \cite{ciuffo2021requiem,gunter2020commercially}. 
For example, the Adaptive Time Gap model \cite{tordeux2010adaptive}, where the time gap $T_n(t)=\big[\Delta x_n(t)-\ell\big]/\dot x_n(t)$, which is the distance gap divided by the velocity, is relaxed to the desired time gap parameter $T$, i.e, $\dot T_n(t)=\frac1\tau[T-T_n(t)]$, or again in Newtonian notation
\begin{equation}
    \dot x_n(t)=\frac{1}{T_n(t)}\Big[\frac{\dot x_n(t)}{\tau}(T_n(t)-T)+\Delta \dot x_n(t)\Big],\qquad\tau,T>0,
    \label{eq:ATG}
\end{equation}
is unconditionally stable for all $\tau,T>0$.

\subsection{Pedestrian models}
\label{subsec:Ped-Models}
\begin{figure}[!ht]
\vspace{-2mm}
    \centering
    \small
\begin{tikzpicture}[x=1.2pt,y=1.2pt]
\definecolor{fillColor}{RGB}{255,255,255}
\path[use as bounding box,fill=fillColor,fill opacity=0.00] (0,0) rectangle (204.86,113.81);
\begin{scope}
\path[clip] (  0.00,  0.00) rectangle (204.86,113.81);
\definecolor{drawColor}{RGB}{0,0,0}

\path[draw=drawColor,line width= 0.4pt,dash pattern=on 1pt off 3pt ,line join=round,line cap=round] ( 42.08, 52.51) -- ( 16.21, 52.51);

\path[draw=drawColor,line width= 0.4pt,dash pattern=on 1pt off 3pt ,line join=round,line cap=round] ( 42.08,  8.61) -- ( 16.21,  8.61);

\path[draw=drawColor,line width= 0.4pt,line join=round,line cap=round] ( 16.21, 52.51) -- ( 16.21,  8.61);

\path[draw=drawColor,line width= 0.4pt,line join=round,line cap=round] ( 19.82, 52.51) --
	( 16.21, 52.51) --
	( 12.60, 52.51);

\path[draw=drawColor,line width= 0.4pt,line join=round,line cap=round] ( 12.60,  8.61) --
	( 16.21,  8.61) --
	( 19.82,  8.61);

\node[text=drawColor,anchor=base,inner sep=0pt, outer sep=0pt, scale=  1.00] at ( 11.04, 28.06) {$\ell$};
\definecolor{drawColor}{gray}{0.40}
\definecolor{fillColor}{gray}{0.92}

\path[draw=drawColor,line width= 2.8pt,line join=round,line cap=round,fill=fillColor] ( 62.60, 30.56) --
	( 62.49, 32.65) --
	( 62.19, 34.71) --
	( 61.68, 36.74) --
	( 60.98, 38.70) --
	( 60.08, 40.58) --
	( 59.01, 42.36) --
	( 57.77, 44.02) --
	( 56.37, 45.55) --
	( 54.83, 46.93) --
	( 53.16, 48.15) --
	( 51.38, 49.19) --
	( 49.51, 50.04) --
	( 47.56, 50.70) --
	( 45.56, 51.16) --
	( 43.53, 51.41) --
	( 41.48, 51.45) --
	( 39.43, 51.29) --
	( 37.41, 50.91) --
	( 35.44, 50.34) --
	( 33.54, 49.57) --
	( 31.72, 48.60) --
	( 30.00, 47.46) --
	( 28.40, 46.15) --
	( 26.94, 44.68) --
	( 25.64, 43.07) --
	( 24.49, 41.33) --
	( 23.52, 39.49) --
	( 22.74, 37.56) --
	( 22.15, 35.56) --
	( 21.76, 33.51) --
	( 21.57, 31.43) --
	( 21.59, 29.34) --
	( 21.81, 27.26) --
	( 22.24, 25.22) --
	( 22.86, 23.23) --
	( 23.67, 21.31) --
	( 24.67, 19.49) --
	( 25.84, 17.77) --
	( 27.18, 16.19) --
	( 28.66, 14.74) --
	( 30.28, 13.46) --
	( 32.02, 12.34) --
	( 33.85, 11.41) --
	( 35.77, 10.67) --
	( 37.75, 10.13) --
	( 39.77,  9.79) --
	( 41.82,  9.66) --
	( 43.87,  9.74) --
	( 45.90, 10.03) --
	( 47.90, 10.52) --
	( 49.83, 11.21) --
	( 51.69, 12.10) --
	( 53.45, 13.17) --
	( 55.10, 14.41) --
	( 56.62, 15.81) --
	( 57.99, 17.37) --
	( 59.20, 19.05) --
	( 60.25, 20.85) --
	( 61.11, 22.75) --
	( 61.78, 24.72) --
	( 62.25, 26.75) --
	( 62.53, 28.82) --
	( 62.59, 30.91) --
	cycle;
\definecolor{fillColor}{RGB}{0,0,0}

\path[fill=fillColor] ( 42.08, 30.56) circle (  3.38);
\definecolor{fillColor}{gray}{0.92}

\path[draw=drawColor,line width= 2.8pt,line join=round,line cap=round,fill=fillColor] (105.71, 83.25) --
	(105.60, 85.34) --
	(105.30, 87.40) --
	(104.79, 89.43) --
	(104.09, 91.39) --
	(103.19, 93.27) --
	(102.12, 95.05) --
	(100.88, 96.72) --
	( 99.48, 98.24) --
	( 97.94, 99.62) --
	( 96.27,100.84) --
	( 94.49,101.88) --
	( 92.62,102.73) --
	( 90.68,103.39) --
	( 88.67,103.85) --
	( 86.64,104.10) --
	( 84.59,104.14) --
	( 82.54,103.98) --
	( 80.52,103.60) --
	( 78.55,103.03) --
	( 76.65,102.26) --
	( 74.83,101.29) --
	( 73.11,100.15) --
	( 71.51, 98.84) --
	( 70.05, 97.37) --
	( 68.75, 95.76) --
	( 67.60, 94.02) --
	( 66.63, 92.18) --
	( 65.85, 90.25) --
	( 65.26, 88.25) --
	( 64.87, 86.20) --
	( 64.68, 84.12) --
	( 64.70, 82.03) --
	( 64.92, 79.95) --
	( 65.35, 77.91) --
	( 65.97, 75.92) --
	( 66.78, 74.00) --
	( 67.78, 72.18) --
	( 68.95, 70.46) --
	( 70.29, 68.88) --
	( 71.77, 67.43) --
	( 73.39, 66.15) --
	( 75.13, 65.03) --
	( 76.96, 64.10) --
	( 78.88, 63.36) --
	( 80.86, 62.82) --
	( 82.88, 62.48) --
	( 84.93, 62.35) --
	( 86.98, 62.43) --
	( 89.01, 62.72) --
	( 91.01, 63.21) --
	( 92.94, 63.90) --
	( 94.80, 64.79) --
	( 96.56, 65.86) --
	( 98.21, 67.10) --
	( 99.73, 68.50) --
	(101.10, 70.06) --
	(102.31, 71.74) --
	(103.36, 73.54) --
	(104.22, 75.44) --
	(104.89, 77.41) --
	(105.36, 79.44) --
	(105.64, 81.51) --
	(105.70, 83.60) --
	cycle;
\definecolor{fillColor}{RGB}{0,0,0}

\path[fill=fillColor] ( 85.19, 83.25) circle (  3.38);
\definecolor{fillColor}{gray}{0.92}

\path[draw=drawColor,line width= 2.8pt,line join=round,line cap=round,fill=fillColor] (183.30, 63.05) --
	(183.20, 65.14) --
	(182.90, 67.20) --
	(182.39, 69.23) --
	(181.68, 71.19) --
	(180.79, 73.07) --
	(179.72, 74.85) --
	(178.48, 76.52) --
	(177.08, 78.05) --
	(175.54, 79.42) --
	(173.87, 80.64) --
	(172.09, 81.68) --
	(170.22, 82.53) --
	(168.27, 83.19) --
	(166.27, 83.65) --
	(164.24, 83.90) --
	(162.19, 83.94) --
	(160.14, 83.78) --
	(158.12, 83.41) --
	(156.15, 82.83) --
	(154.24, 82.06) --
	(152.42, 81.09) --
	(150.71, 79.95) --
	(149.11, 78.64) --
	(147.65, 77.17) --
	(146.34, 75.56) --
	(145.20, 73.83) --
	(144.23, 71.99) --
	(143.45, 70.05) --
	(142.86, 68.05) --
	(142.47, 66.00) --
	(142.28, 63.92) --
	(142.30, 61.83) --
	(142.52, 59.76) --
	(142.95, 57.71) --
	(143.57, 55.72) --
	(144.38, 53.80) --
	(145.38, 51.98) --
	(146.55, 50.26) --
	(147.89, 48.68) --
	(149.37, 47.24) --
	(150.99, 45.95) --
	(152.72, 44.84) --
	(154.56, 43.90) --
	(156.48, 43.16) --
	(158.46, 42.62) --
	(160.48, 42.28) --
	(162.53, 42.15) --
	(164.58, 42.23) --
	(166.61, 42.52) --
	(168.61, 43.01) --
	(170.54, 43.70) --
	(172.40, 44.59) --
	(174.16, 45.66) --
	(175.81, 46.90) --
	(177.33, 48.31) --
	(178.70, 49.86) --
	(179.91, 51.54) --
	(180.96, 53.34) --
	(181.82, 55.24) --
	(182.49, 57.21) --
	(182.96, 59.25) --
	(183.23, 61.32) --
	(183.30, 63.40) --
	cycle;
\definecolor{fillColor}{RGB}{0,0,0}

\path[fill=fillColor] (162.78, 63.05) circle (  3.38);
\definecolor{drawColor}{RGB}{0,0,0}

\path[draw=drawColor,line width= 0.4pt,dash pattern=on 4pt off 4pt ,line join=round,line cap=round] ( 42.08, 30.56) -- ( 85.19, 83.25);

\path[draw=drawColor,line width= 0.4pt,dash pattern=on 4pt off 4pt ,line join=round,line cap=round] ( 42.08, 30.56) -- (162.78, 63.05);

\path[draw=drawColor,line width= 0.8pt,line join=round,line cap=round] (110.62, 44.46) -- ( 98.12, 41.10);

\path[draw=drawColor,line width= 0.8pt,line join=round,line cap=round] (101.69, 44.68) --
	( 98.12, 41.10) --
	(103.01, 39.79);

\node[text=drawColor,anchor=base,inner sep=0pt, outer sep=0pt, scale=  1.00] at (110.19, 34.21) {$\vec e_{n,p}$};

\path[draw=drawColor,line width= 1.6pt,line join=round,line cap=round] ( 42.08, 30.56) -- ( 84.33, 21.84);

\path[draw=drawColor,line width= 1.6pt,line join=round,line cap=round] ( 75.41, 18.88) --
	( 84.33, 21.84) --
	( 77.31, 28.08);

\node[text=drawColor,anchor=base,inner sep=0pt, outer sep=0pt, scale=  1.00] at ( 67.94, 15) {$\dot{\vec x}_n$};

\node[text=drawColor,anchor=base,inner sep=0pt, outer sep=0pt, scale=  1.00] at ( 42.94, 19.04) {$\vec x_n$};

\node[text=drawColor,anchor=base,inner sep=0pt, outer sep=0pt, scale=  1.00] at ( 86.05, 90.53) {$\vec x_m$};

\node[text=drawColor,anchor=base,inner sep=0pt, outer sep=0pt, scale=  1.00] at (164.51, 71.53) {$\vec x_p$};

\path[fill=fillColor] ( 42.08, 30.56) circle (  3.38);
\definecolor{drawColor}{gray}{0.60}

\path[draw=drawColor,line width= 1.2pt,line join=round,line cap=round] ( 42.08, 30.56) -- ( 85.19, 83.25);

\path[draw=drawColor,line width= 1.2pt,line join=round,line cap=round] ( 38.16, 33.76) --
	( 42.08, 30.56) --
	( 45.99, 27.36);

\path[draw=drawColor,line width= 1.2pt,line join=round,line cap=round] ( 89.10, 80.05) --
	( 85.19, 83.25) --
	( 81.27, 86.45);
\definecolor{drawColor}{RGB}{0,0,0}

\node[text=drawColor,anchor=base,inner sep=0pt, outer sep=0pt, scale=  1.00] at ( 50.60, 62.11) {$|\Delta \vec x_{n,m}|$};
\end{scope}
\end{tikzpicture}
    \caption{Main variables of pedestrian dynamics in two dimensions. $d_{n,m}$ is the distance from pedestrian $n$ to pedestrian $m$, while $\vec e_{n,p}$ is the unit vector joining pedestrian $p$ to pedestrian $n$.}
    \label{fig:scheme_Ped}
\end{figure}
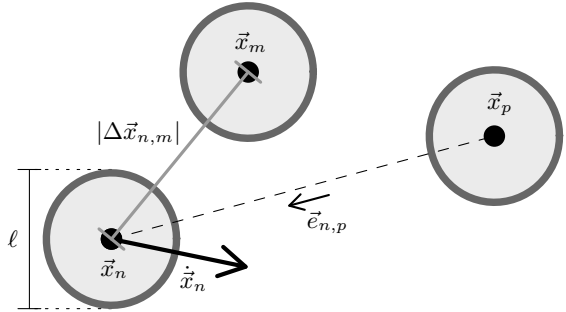

Turning to pedestrians, the world gains a new geometric dimension, as pedestrians do not evolve on predefined tracks. Thus, pedestrian models are two-dimensional models that describe the dynamics of the pedestrian position $\vec x_n(t)\in\mathbb R^2$ according to the positions and velocities of its neighbours. 
In contrast to car-following models which have nearest neighbour interaction schemes in one dimension, the modelled pedestrians typically interact with all their close neighbours in two dimensions. 
The interactions may depend on distance, velocity difference, and also safety related variables such as the time-to-collision (TTC). 
Pioneering models are based on a superposition of pairwise interaction forces with the neighbours, the so-called \emph{social forces} \cite{helbing1995social}, while more recent approaches rely on collision avoidance, anticipation, and optimisation concepts. 
Pedestrians are generally represented as disks, as in Figure~\ref{fig:scheme_Ped}, although more realistic descriptions use ellipses or even skeleton models for the pedestrian shapes.

\subsubsection{Force-based models}

The concept of a \emph{social force field} can be traced back to the seminal works of Lewin in the social psychology of group dynamics \cite{lewin1952field,lewin1968conceptual,chraibi2011force}. 
\emph{Force-based} pedestrian models date back to the pioneering work of Hirai and Tarui in the 1970s on the simulation of crowd behaviour in panic \cite{hirai1975simulation} where the dynamics of a pedestrian result from the superposition of different pairwise forces, including \emph{attraction and repulsion forces} with the neighbours and \emph{velocity alignment force}.
The concept of \emph{social force} for pedestrian dynamics was later democratised from the 1990s with the work of Helbing and co-authors \cite{helbing1995social,helbing2000simulating}. 
Around the same time, swarming-type models such as the \emph{Boids model} by Reynolds \cite{reynolds1987flocks} and \emph{self-propelled particle model} by Vicsek and co-authors \cite{vicsek1995novel}, or, later, Cucker-Smale and other swarming and flock models \cite{cucker2007emergent,carrillo2010particle} have been developed with interaction mechanisms based on averages with neighbours and similar principles of pairwise interaction superposition.
However, in contrast to swarming models, space is not isotropic in pedestrian models: agents experience a drift towards their desired destinations in addition to the interaction field.

Minimal force-based models are given by the second-order dynamics
\begin{equation}
    \ddot{\vec x}_n(t)=\frac1\tau\big[\vec v_0-\dot{\vec x}_n(t)\big]+\sum_{m\ne n}
    \omega\big(\phi_{n,m}(t)\big)\,
    \nabla_{\!\vec x\;}\mathcal U\Big[\Delta \vec x_{n,m}(t),\Delta \dot{\vec x}_{n,m}(t)\Big],
    \label{eq:FB}
\end{equation}
where $\tau>0$ is a reaction (relaxation) time, $\vec v_0\in\mathbb R^2$ is the \emph{desired velocity} coming from a tactical modelling level, the social force is the gradient in space of the interaction potential $\mathcal U\in C^2(\mathbb R^4,\mathbb R^2)$, while $\omega\in C^1(\mathbb R^2,\mathbb R)$ is an anisotropic factor based on the bearing angle $\phi_{n,m}$ that provides more weight to the pedestrians in the direction of motion.
For example, in the pioneering \emph{Social Force} (SF) model, the interaction force is derived from an exponential repulsive distance-based interaction potential and a piecewise constant anisotropic factor
\begin{equation}
    \mathcal U_{SF}(\Delta \vec x)=AB\exp(-|\Delta \vec x|/B),\quad \omega_{SF}(\phi)=\left\{\begin{array}{cc}
        1 & \text{if }|\phi|<\kappa, \\
        c & \text{otherwise},
    \end{array}\right.
\end{equation}
where $A,B>0$ are the repulsion range and the characteristic distance, respectively, while $0<c \ll 1$.
Due to the exponential potential, the interaction of the social force model is short-ranged. 

In the \emph{Centrifugal Force} (CF) model \cite{yu2005centrifugal} and the \emph{Generalised Centrifugal Force} (GCF) model \cite{chraibi2010generalized}, the range of interaction decays algebraically. 
The repulsive force for the CF model is
\begin{equation}
    \nabla_{\!\vec x\;}\mathcal U_{CF}(\Delta \vec x,\tilde\Delta \dot{\vec x})=\beta\frac{|\tilde\Delta \dot{\vec x}|^2}{|\Delta \vec x|}\frac{\Delta \vec x}{|\Delta \vec x|},
\end{equation}
while the force for the GCF model reads
\begin{equation}
    \nabla_{\!\vec x\;}\mathcal U_{GCF}(\Delta \vec x,\tilde\Delta \dot{\vec x})=\beta\frac{\big[\eta\vec v_0+\tilde\Delta \dot{\vec x}\big]^2}{|\Delta \vec x|+r(\dot{\vec x})}\frac{\Delta \vec x}{|\Delta \vec x|}, 
\end{equation}
where $\beta,\eta>0$ are parameters, $r\in C(\mathbb R^2,R_+)$ models a velocity-dependent pedestrian shape, e.g. an ellipse \cite{chraibi2010generalized}, while $\tilde\Delta \dot{\vec x}$ is the relative velocity projected on the direction to the neighbouring pedestrians. 
Unlike the SF model, the CF and GCF models also rely on the relative velocity to the neighbours. 
Nowadays, many extensions of the SF model include relative velocity terms and other velocity-dependent mechanisms; see the review \cite{chen2018social} and references therein. 
For instance, in \cite{parisi2009modification}, the SF model extension consists of some respect areas in the vicinity of the pedestrians that force them to stop. 
Similarly, but in a dynamic way,  in \cite{zanlungo2011social} the SF model is extended with explicit collision prediction. 
In another recent approach, the interaction force is anisotropic to take into account for a preferred crossing direction \cite{totzeck2020anisotropic}.

\subsubsection{Velocity-based models}
For pedestrian models, first-order models came second to second-order ones: 
A stream of models developed subsequently to the force-based models are the velocity-based models. 
In contrast to their Newton-like forebears, velocity-based models instantly regulate velocity at the first order. 
They are inherently free of inertia and allow strict control of body exclusion - an aspect that is difficult to manage with a force-based model \cite{chraibi2011force,sticco2020effects}. 
Velocity-based models are initially borrowed from the robotics.
They generally rely on velocity optimisation processes under collision constraints over given anticipation times. 

Such developments can be traced back to the pioneering work of Fiorini et al.\ using purely geometric \emph{Velocity Obstacle} (VO) models \cite{fiorini1998motion}. 
The model is based on linear extrapolations of the pedestrian trajectories to determine potential collision sets. Geometrically, these sets assume the form of so called collision cones under the hypothesis of constant pedestrian velocities. The cone is empty if no collision is expected.  
The velocity dynamics of agent $n$ is then obtained by minimising the deviation from the desired velocity under the exclusion constraint of the time-dependent collision cones $\mathcal{C}_n^m(t)$ with other agents $m$
\begin{equation}
    \dot{\vec x}_n(t)=\text{arg}\hspace{-5mm}\min_{\vec v_n\,\not\in\,\cup_{m\ne n}\mathcal{C}_n^m(t)}\|\vec v_0-\vec v_n\|^2
    \label{eq:VO}
\end{equation}
where $\vec v_0$ is the desired velocity. 
Since the collision cone depends on the current velocities of the agents, the velocity model is implicit. 
In practice, the implicit system is solved numerically using semi-implicit schemes combining an explicit Euler solver for the velocity with an implicit Euler scheme for the positions of the agents. 
Initially, velocity obstacle models were developed for controlling robots in an environment including moving obstacles and automated vehicles driving on a highway \cite{fiorini1998motion}.
First applications for pedestrian dynamics date back to the late 2000s \cite{paris2007pedestrian,van2008reciprocal}.

Generally speaking, velocity obstacle models can be formulated as an optimisation problem under constraints \cite{van2020generalized}
\begin{equation}
    \dot{\vec x}_n(t)=\text{arg}\hspace{-5mm}\min_{\vec v_n\,\not\in\,\cup_{m\ne n}\mathcal{C}_n^m(t)}C_{\vec x(t)}^{\vec v_0}(\vec v_n)
\end{equation}
where $C_{\vec x}^{\vec v_0}$ is a cost function depending on the desired velocity and also the positions and velocities of the neighbours.
In the collision-free model developed in \cite{paris2007pedestrian}, the cost function includes components associated with speed variations, distance to desired speed, and orientation changes. 
In the \emph{Reciprocal Velocity Obstacle} (RVO) model \cite{van2008reciprocal}, the cost is the sum of a quadratic deviation from the desired velocity as in the original velocity obstacle model plus a repulsive potential based on the time to collision. 
The cost function is a quadratic deviation from the desired speed as in the initial VO model \eqref{eq:VO} in the \emph{Optimal Reciprocal Collision Avoidance} (ORCA) model \cite{van2011reciprocal}, while its extension PORCA model \cite{luo2018porca} includes an additional velocity-dependent term that prevents the system from freezing.
No collision cones are used in the \emph{Linear Trajectory Avoidance} model \cite{pellegrini2009you}. 
Instead, the optimisation is done by exponentially penalising the minimum distance under linear extrapolation to avoid collisions, in addition to a quadratic deviation from the desired speed as in the original VO model \eqref{eq:VO}.

A general collision-free velocity-based modelling framework was proposed in the 2000s by Maury and Venel \cite{maury2008mathematical,maury2011discrete}. 
Its principle also rests on velocity optimisation under collision-free constraints.
However, unlike the models described above where the pedestrian velocities are optimised individually, the optimisation is performed centrally over all the pedestrians. 
The pedestrians are considered as disks of diameter $\ell$.
The set of admissible configurations of the system excluding overlapping configurations is given by
\begin{equation}
    Q=\{\vec x,~ D_{m,n}=|\Delta \vec x_{n,m}|-\ell\ge0,~~\forall m\ne n\}.
\end{equation}
Then the set of admissible velocities constrains the system to remain in the admissible configurations
\begin{equation}
    C_{\vec x}=\{\dot{\vec x},~ D_{m,n}=0\Rightarrow \nabla D_{m,n}\cdot\dot{\vec x}_n\ge0,~~\forall m\ne n\}.
\end{equation}
If a contact occurs, i.e., $D_{m,n}=0$, then the admissible velocity enforces the motion away as $\nabla D_{m,n}\cdot\dot{\vec x_n}\ge0$ and so the distance increases.
This formally requires the system to remain within the admissible configurations. 
Finally, similar to velocity obstacle models \eqref{eq:VO}, the velocities are computed using the orthogonal projection of the desired velocity field onto the admissible velocities
\begin{equation}
    \dot{\vec x}(t)=P_{C_{\vec x(t)}}\vec v_0.
\end{equation}
The use of orthogonal projection models dynamics where pedestrians try to reach their destination as quickly as possible under collision-free constraints, as in an emergency evacuation.

\subsubsection{Hybrid models}

Many pedestrian models are hybrid approaches combining different concepts and mechanisms from first and second order models such as social force, optimal velocity (OV), or optimisation and collision avoidance techniques. 
For example, some extensions of force-based models are inspired by swarming approaches \cite{helbing2000simulating,lakoba2005modifications}, where the desired velocity $\vec v_0$ in \eqref{eq:FB} is mitigated with the average velocity of the neighbouring pedestrians using
\begin{equation}
    \underline{\vec v}_0(t)= p \vec v_0+(1-p)\langle \dot x_m(t)\rangle_n,\qquad p\in[0,1],
\end{equation}
to address mass collective behaviour. 
In the same vein, the pedestrian velocity is relaxed to a combination of the desired velocity and a superposition of distance-based OV models \cite{nakayama2005instability,nakayama2008effect}, generalising 
concepts of car-following models \cite{bando1995dynamical,bando1998analysis,jiang2001full} to two dimensions
\begin{equation}
    \ddot{\vec x}_n(t)=\frac1\tau\Big[\big[\vec v_0+\sum_{m\ne n} f\big(|\Delta \vec x_{n,m}(t)|\big)\omega\big(\phi_{n,m}(t)\big)\vec e_{n,m}(t)\big]-\dot{\vec x}_n(t)\Big],
    \label{eq:PedOVM}
\end{equation}
where $f(x)=\alpha\big(a+\text{atan}\beta(x-b)\big)$ is the OV function, 
$\omega(x)=1+\cos(x)$ is an anisotropic vision field factor, and $\vec e_{n,m}(t)$ is the unit vector directed from $n$ to $m$. 

Another hybrid two-dimensional OV pedestrian model relies on first-order dynamics \cite{tordeux2016collision}.
As in the modelling framework introduced by Maury and Venel in \cite{maury2008mathematical,maury2011discrete}, this model constrains pedestrian velocities to remain in admissible collision-free configurations \cite{tordeux2016collision}.
However, the speed is regulated according to the distance to the next pedestrian in the direction of motion using an OV function, while the direction results from a superposition of distance-based repulsion with the neighbours as in force-based models.
An extension of this collision-free velocity-based model relaxes the direction, making it a second-order model \cite{xu2019generalized}. 
In the \emph{Gradient Navigation} model introduced in \cite{dietrich2014gradient}, the speed is relaxed to the OV function, while the direction results from a the desired velocity and a superposition of distance-based repulsion with the neighbours as in force-based models
\begin{equation}
    \left\{\begin{array}{l}
         \dot{\vec x}_n(t) = w_n(t)\big[C_1\vec v_0+C_2\sum_{m\ne n} \nabla\mathcal U_{n,m}\big],  \\[2mm]
         \dot w_n(t)= \frac1\tau\big[V(\rho_n)-w_n(t)\big], 
    \end{array}\right.
\end{equation}
where $C_1$ and $C_2$ are two normalisation constants while $V(\rho_n)$ is the OV depending on the local density $\rho_n$.
A similar approach is given in \cite{moussaid2011simple} with an heuristic model without social interaction and where, as in the collision-free velocity-based model \cite{tordeux2016collision}, the OV depends on the distance to the nearest pedestrian in front.

The concept of anticipatory collision avoidance in force-based models introduced in \cite{zanlungo2011social} is deepened in a recent hybrid force-based model \cite{karamouzas2014universal}, where the interaction force is derived directly from the potential
\begin{equation}
    \mathcal U_{TTC}(\tau_{n,m})=\frac{a}{\tau_{n,m}^2}\exp(-\tau_{n,m}/\tau_0),\qquad a,\tau_0>0,
    \label{eq:PotTTC}
\end{equation}
based on the time-to-collision (TTC)
\begin{equation}
    \tau_{n,m}=\left\{\begin{array}{ll}
         \frac{\Delta\vec x_{n,m}\Delta\dot{\vec x}_{n,m}-\sqrt{\Delta}}{\Delta\dot{\vec x}_{n,m}^2}& \text{if }\Delta>0, \\[1mm]
         \infty&\text{otherwise}, 
    \end{array}\right.~~
    \Delta=\big(\Delta\vec x_{n,m}\cdot\Delta\dot{\vec x}_{n,m}\big)^2-|\Delta\dot{\vec x}_{n,m}|^2(|\Delta\vec x_{n,m}|^2-\ell^2).
    \label{eq:TTC}
\end{equation}
The TTC $\tau_{n,m}$ is the time before a collision occurs between the $n$-th and $m$-th pedestrians (i.e., $|\Delta\dot{\vec x}_{n,m}|<\ell$), assuming that their velocities are constant.
In \cite{lu2020pedestrian}, the force is based on similar collision anticipation mechanisms.
A general hybrid modelling framework first introduced in the early 2000s \cite{hoogendoorn2003simulation} assumes pedestrian dynamics based on a \emph{decisional layer}, resulting from an optimal control problem, and a \emph{mechanical layer} modelling physical exclusion force and other inertia phenomena
\begin{equation}
    \tilde{\vec v}_0^n(t)=\text{arg}\hspace{-1mm}\min_{\vec v_n\in\mathbb R^2}\mathcal E_{\vec x(t)}^{\vec v_0}(\vec v_n),\qquad \ddot{\vec x}_n(t)=\frac1\tau\big[\tilde{\vec v}_0^n(t)-\dot{\vec x}_n(t)\big]+F^\text{M}_n(t),
\end{equation}
where $\mathcal E$ is an energy function involving several concepts, while $F^\text{M}_n(t)$ are the mechanical force on the $n$-th pedestrian at time $t$. 
In the \emph{ANticipatory Dynamic Algorithms} (ANDA) \cite{echeverria2023body}, the energy function consists of four main components:
\begin{enumerate}
    \item \emph{Target energy}, which models the desired destination and velocity from a tactical modelling level,
    \item \emph{Speed energy} including bio-mechanical costs for walking speed and inertia,
    \item \emph{Repulsion energy} in the form of a superposition of pairwise nonlinear distance-based repulsive potentials with the neighbours,
    \item \emph{Anticipation energy} based on the TTC potential \eqref{eq:PotTTC}-\eqref{eq:TTC} introduced in \cite{karamouzas2014universal}.
\end{enumerate}

\section{Modelling collective dynamics}
\label{sec:CollDyn}

Traffic and pedestrian flows describe many coordinated collective dynamics \cite{helbing2001traffic,chowdhury2000statistical,castellano2009statistical,boltes2018empirical}
A typical example in traffic flow is stop-and-go dynamics on highways where jam waves resulting from instability phenomena \cite{orosz2010traffic,wilson2011car}, also called \emph{jamitons} \cite{flynn2009self}, propagate upstream in the flow without any apparent external cause. 
Typical examples for pedestrian dynamics are lane formation for counterflow, jamming and clogging at bottlenecks, oscillation and flow intermittency for counterflow at bottlenecks, or stripe formation for crossing flow \cite{helbing2005self,boltes2018empirical}.
Most of the aforementioned models describe these phenomena, with realistic features for fine-tuning the parameters.  
However, conventional models fall short of replicating more complex situations involving long-term anticipation (on the order of several seconds) or socio-psychological aspects. 
A typical example is the formation of a clear lane in pedestrian crowds to facilitate the passage of a vehicle, typically an ambulance \cite{MadrasAmbulance}, which relies on early anticipation. 
Another example in the load balancing of pedestrians over several exits in case of normal evacuation. 
Here, the distance between the exits requires pedestrians to anticipate for long periods of time.
These different collective phenomena and related modelling approaches are discussed in more detail in the following sections.

\subsection{Stop-and-go waves in traffic flow}

In this section, we focus on stop-and-go dynamics in congested traffic and show how collective linear stability analysis can be used to address this phenomenon.

\subsubsection{Empirics of stop-and-go waves in traffic flow}

Stop-and-go dynamics are currently observed in road traffic flow. 
They are also found in single-file pedestrian movements and in bicycle flow \cite{zhang2014universal}.
In fact, stop-and-go and accordion traffic  now commonly denotes congested traffic. 
In addition to its scientific interest, the jerky dynamics of congested traffic is a major safety concern.
It also has a negative impact on the environment. 
Indeed, the acceleration and deceleration phases of stop-and-go traffic dynamics cause an increase in fuel consumption and pollutant emissions \cite{li2014stop}.

\begin{figure}[!ht]
    \centering
    \includegraphics[width=.75\textwidth]{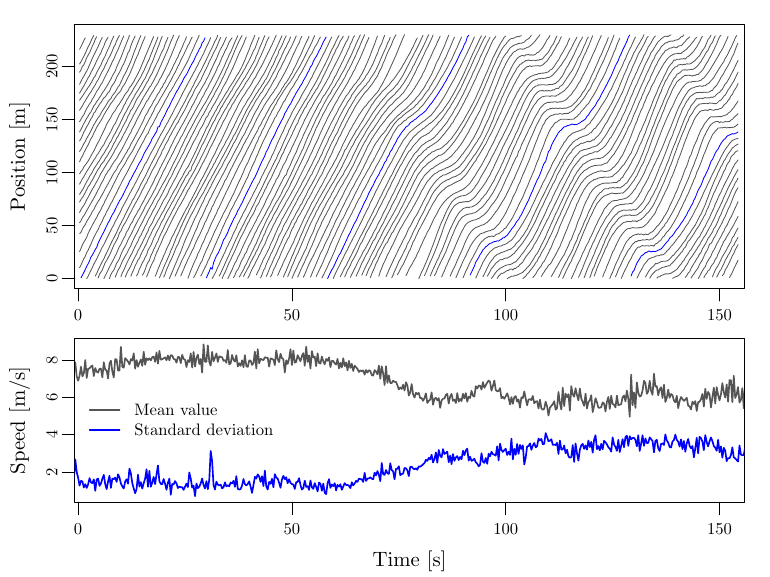}
    \caption{Trajectories in the experiment of Sugiyama et al.\ with 22 vehicles on a circuit of length 230 meters starting from an uniform configuration \cite{sugiyama2008traffic}. After a while, a stop-and-go wave appears, which causes a decrease in the average speed and an increase in the speed standard deviation.}
    \label{fig:StopAndGo}
\end{figure}

The terminology of \emph{stop-and-go waves} can be traced back to the late 1960s \cite{duckstein1967control}. 
A related word is \emph{phantom jam} \cite{sugiyama2008traffic}, as the waves seem to appear for no apparent reason. 
This was demonstrated in the late 2000s with an experiment involving 22 vehicles on a 230-metre circuit \cite{sugiyama2008traffic} (see Fig.~\ref{fig:StopAndGo}). 
Starting from a homogeneous configuration with equal distance gaps, a stop-and-go wave emerged after a while, with jerky vehicular dynamics not caused by any infrastructure peculiarity. 
This experiment shed new light on the formation of waves in traffic flow, and has helped to popularise it.
A more recent experiment under similar conditions has shown that a single vehicle with particularly stable dynamics allows the waves to dissipate.
This field of research remains particularly active (see section~\ref{sec:Challenges} below).

\subsubsection{Collective stability analysis}

The formation of stop-and-go waves and phase separation from laminar traffic to oscillatory dynamics was theorised in the 2010s, mostly using linear stability analysis \cite{treiber2013traffic,orosz2010traffic,wilson2011car}.
Local stability analysis considers a platoon of vehicles following a leader. 
Stability conditions are usually derived using Laplace transform and transfer function. 
Global (or string or collective) stability analysis considers a flow of vehicles with periodic boundary conditions. 
In general, global stability conditions are more restrictive than local ones, since global analysis takes into account advective or convective perturbations that vanish locally. 
The stability conditions for infinite systems are even more restrictive. 
The stability of finite periodic systems can be derived using matrix spectral analysis, while for infinite systems an exponential ansatz is generally used. The latter is equivalent to performing a Laplace transformation in time and a discrete Fourier transformation in space \cite{orosz2010traffic}.

To be more specific, consider the linearisation of the car-following model around the homogeneous equilibrium solution  
\begin{equation}
    \ddot{\tilde x}_n(t) = a\Delta\tilde x_n(t)) + b \dot{\tilde x}_n(t) + c \dot{\tilde x}_{n+1}(t)
\end{equation}
and $N$ vehicles with periodic boundary conditions on a circuit of lane $L$ (i.e., the $N$-th vehicle interacts with the first vehicle and its spacing is $\Delta x_N(t)=x_N(t)-x_1(t)+L$). 
The matrix system being circulant, it is easy to check that the characteristic equation of the dynamics reads
\begin{equation}
    \lambda^2+\lambda(b+c\iota_k)+a(1-\iota_k)=0,\quad \iota_k=\exp\big(\mathrm i 2\pi k/N\big),~k=0,\ldots,N-1.
\end{equation}
Here $\iota_k$ is the $k$th root of the unit, which appears thanks to the circulatory property of the linear system.
The equation is a second order complex polynomial or again a real polynomial of order $2N$.
Here $\lambda=0$ is a solution for $k=0$ resulting from the periodic boundary where the sum of the spacings $\sum_{n=1}^N\Delta x_n(t)=L$ is conserved at any time, while the real part of the remaining eigenvalues are strictly negative (i.e., the system is linearly stable) if 
\begin{equation}
    a>0,\qquad b<0\qquad\text{and}\qquad b^2-c^2>2a,
\end{equation}
see, e.g., \cite{tordeux2012linear}, \cite[Chap.~15]{treiber2013traffic} and general stability conditions for complex polynomials in \cite[Th.~3.2]{Frank1946OnTZ} which are a generalisation of the Hurwitz conditions. 
The conditions are sufficient for any finite system and become exact for infinite systems.
For example, the linear stability condition is 
\begin{equation}
    0<2\tau<1/V'(L/N)
\end{equation}
for the OV model~\eqref{eq:OVM},
\begin{equation}
    0<2\tau_1\tau_2/\big(2\tau_1+\tau_2\big)<1/V'(L/N)
\end{equation}
for the FVD model~\eqref{eq:FVDM}, while the ATG model~\eqref{eq:ATG} is systematically linearly stable provided that $\tau,T>0$, hence its prevalence for automated driving systems \cite{khound2023extending}.

\subsection{Lane formation in pedestrian dynamics}

In this section, we turn to the collective dynamics of lane formation for counterflow of pedestrians moving in opposite directions and the class of models that describe this phenomenon.

\subsubsection{Empirical observations in pedestrian and other particle systems}

\begin{figure}[b]
    \medskip\centering
    $\qquad$\includegraphics[width=0.55\textwidth]{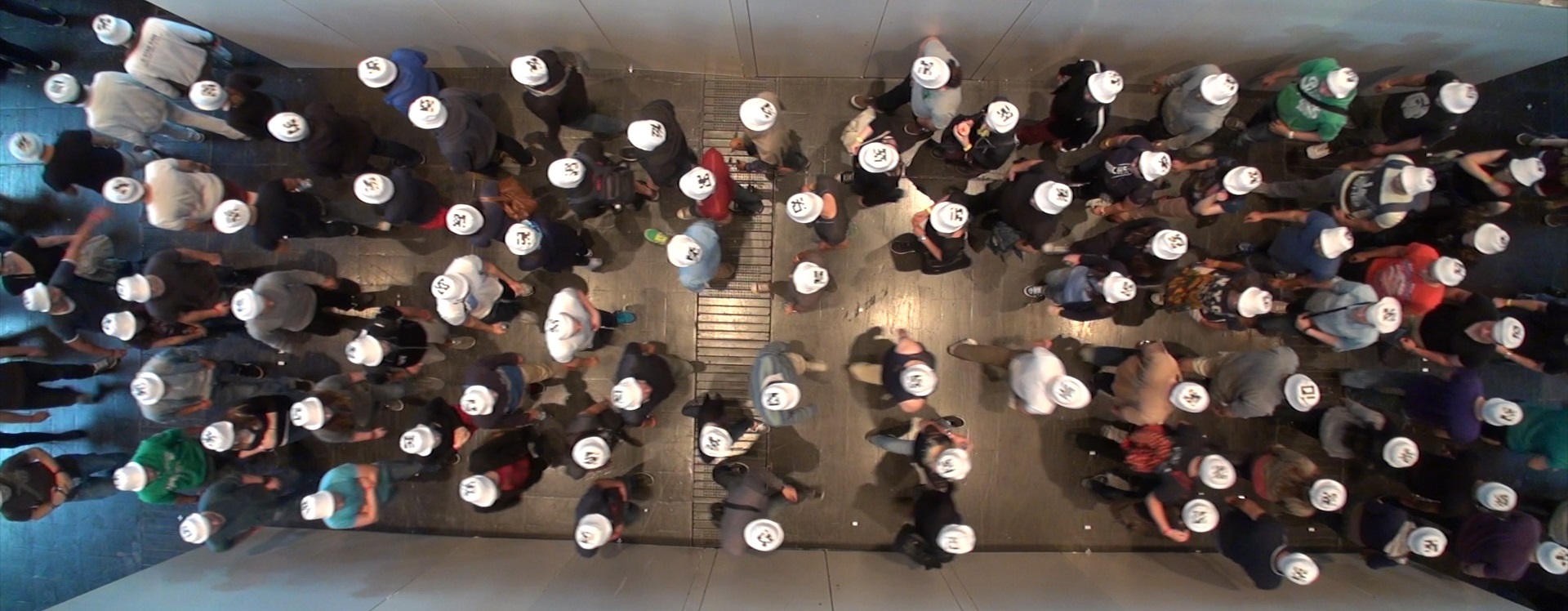}\\[2mm]
    \includegraphics[width=0.75\textwidth]{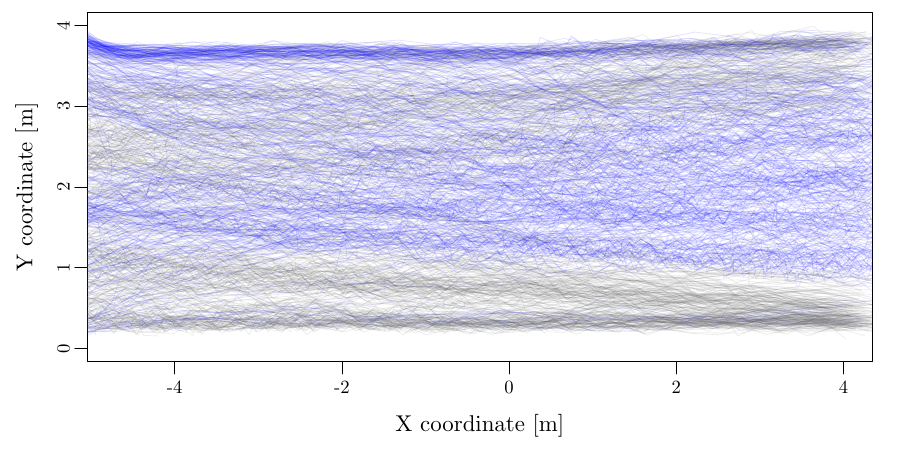}\vspace{-2mm}
    \caption{Trajectories of pedestrians in a counterflow experiment \cite{cao2017fundamental}. The blue trajectories are pedestrians moving from right to left, while the grey trajectories are pedestrians moving from left to right. We can see the formation of lanes moving in opposite directions.}
    \label{fig:LaneFormation}
\end{figure}

Counterflows, where two groups of pedestrians move in opposite directions, often lead to the formation of lanes of pedestrians walking in the same direction. 
Lane formation in pedestrian dynamics has been observed since the 1960s \cite{oeding1963verkehrsbelastung,navin1969pedestrian} and is studied in experiments (see Fig.~\ref{fig:LaneFormation} and \cite{cao2017fundamental}). 
It may also be related to the stripes formed at intersections between pedestrian streams \cite{bacik2023lane}. 
In fact, the phenomenon of lane formation is widely encountered in discrete systems in which entities move in two opposite directions \cite{dzubiella2002lane}. 
For example, lane formation occurs in passive matter, e.g., driven mixtures of oppositely charged colloids \cite{vissers2011lane} or driven plasmas \cite{sutterlin2009dynamics,sarma2020lane}, and in active matter, such as in microswimmers \cite{kogler2015lane}, bacterial populations \cite{shimaya2019lane}, or social insects such as ants \cite{couzin2003self}.
Nowadays, laning phenomenon is recognised as a fundamental collective property of particle and self-avoiding agent systems driven by a mixture of two potentials \cite{cristin2019general,reichhardt2018laning}.

\subsubsection{Modelling lane formation}

Lane formation in pedestrian and other mixed-particle models is difficult to analyse mathematically and is usually studied by simulation or using continuous appraoches (see, e.g., \cite{bacik2023lane}). 
In fact, most of two-dimensional pedestrian models are able to reproduce laning in counterflow. 
Pioneering simulation results were obtained in the 2000s using the social force model \cite{helbing2000simulating}, basic cellular automata \cite{burstedde2001simulation}, or, a few years later, velocity obstacle models \cite{guy2010pledestrians}. 
The formation of lanes can also be explained by socio-psychological factors of group membership and social identity \cite{templeton2020placing,wijermans2022towards}.
In \cite{khelfa2022heterogeneity}, lane formation is even observed for mixed flows moving in the same direction but at different speeds. 
However, some experiments with pedestrians walking at different speeds did not provide empirical evidence of lane formation \cite{fujita2019traffic}. 

In fact, the occurrence of lane formation is observed also in systems of inanimate particles. 
For example in \cite{sutterlin2009dynamics}, the volume exclusion and collisions due to the motion in different directions in driven binary complex plasma, combined with joint pushing of particles moving in the same direction, induce a segregation into lanes. Thus, any minimal models for pedestrian dynamics representing volume exclusion, such as force models with repulsive interactions, are able to reproduce lane formation. This example indicates that lane formation itself does not need social or psychological explanations, even if the steering process of humans in such a situation is based on the impressive cognitive and biomechanical abilities of the human body. 
These aspects are discussed in more detail in \cite{sieben2017collective}.

In general, lane formation in pedestrian counterflow depends on density.
At low densities, it may be advantageous to avoid rather than follow, thus limiting lane formation. 
Conversely, at high densities, exclusion constraints may prevent the formation of lanes.
However, there is no empirical evidence for a transition, although experiments have shown that counterflow can give rise to gridlock at densities greater than 3.5 persons per square metre \cite{zhang2012ordering}. 
In the models, these density thresholds above which lanes no longer emerge depend on the form of the model in addition to the initial configuration \cite{xu2021anticipation,tordeux2018mesoscopic}.
They remain an open rearch question.
Another factor that leads to counterflow gridlock is the introduction of noise into pedestrian motion models. 
This phenomenon is known in the literature as  \emph{freezing by heating} \cite{helbing2000freezing}. 
Even in simple deterministic models, a transition to lane formation occurs as the driven force increases relative to the interacting repulsive forces \cite{tordeux2023multi}.

\subsection{Far-sighted anticipation in local navigation}

Rather than lanes arising from oppositely-directed flows, let us now inspect the response of a static crowd to a moving `obstacle'. For this purpose, we focus on an experiment presented in \cite{nicolas2019mechanical} that highlights the long-term anticipatory abilities of pedestrians to avoid an intruder, and how mean-field games can be used to model this type of behaviour.

\subsubsection{Empirical evidence for long-term anticipation}

That pedestrians or cars plan their motion tens of minutes or even hours ahead should not come as a surprise. Unlike physical particles, their local navigation is steered towards a target destination set at larger modelling scales. The existence of a desired direction, imparted to agents at the higher tactical level, makes space anisotropic for them (i.e., not all directions are equivalent), irrespective of their interactions with the rest of the crowd, in contrast to boids \cite{reynolds1987flocks} or simple swarming models \cite{vicsek1995novel} for which directional preferences only result from interactions.

What may be less obvious is that even the local, operational interactions with their neighbours and the built environment may involve some degree of planning. Direct visualisation of the simulation output of simple force-based models and the frequent near-collisions they imply evince that distance-based interaction forces are not sufficient to render realistic collision avoidance.
Taking into account the relative velocities of the agents is thus necessary. Evaluating distances via this relative velocity, i.e., in terms of the imminence of a collision risk, gives rise to the well-known notion of anticipated times to collision (TTC). Experimental observations (collected in real life or in virtual reality) as well as empirical data show the central role played by the TTC for collision avoidance and the spatial arrangement of crowds \cite{karamouzas2014universal,pfaff2018avoidance,meerhoff2018guided,cordes2024dimensionless}.
Interactions between people may start at a TTC of several seconds, when they are more than half a dozen meters away from each other.

\begin{figure}[!ht]
    \medskip\centering
    \includegraphics[width=0.25\linewidth]{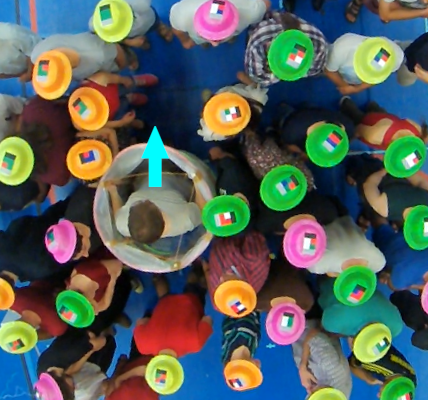} \qquad\qquad
    \includegraphics[width=0.25\linewidth]{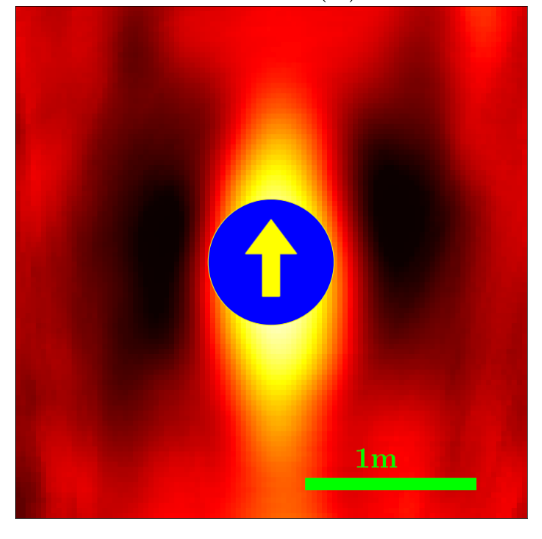}

    \hspace{0.34\linewidth}\includegraphics[width=0.3\linewidth]{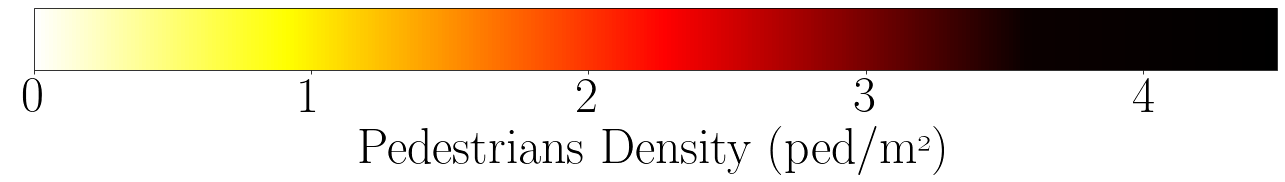}
    \caption{Experiment~\cite{nicolas2019mechanical} in which a dense crowd is crossed by an cylindrical intruder (Left). Pedestrians anticipate the passage of the cylinder as seen from the void in front of the cylinder, and temporarily accept the discomfort of the dense regions on both sides of the cylinder (Right, from~\cite{butano_a_u2024a}), because they know it will not last.}
    \label{fig:intruder}
\end{figure}

Interestingly, anticipation may even need to be pushed beyond the nearest collision to describe operational pedestrian dynamics \cite{wagoum2017understanding}. In particular, a crowd's response to an intruder's crossing~\cite{nicolas2019mechanical} (by transverse moves opening a depleted tunnel ahead of the intruder, see Fig.~\ref{fig:intruder}) appears to elude state-of-the-practice models based (or not) on TTC~\cite{bonnemain2023pedestrians, butano_a_u2024a}. 
The observed response could actually be understood by assuming that pedestrians anticipate much beyond the next collision, and actually beyond the passage of the moving obstacle, as we shall see.

\subsubsection{Mean-field games}

Situations where pedestrians have to compete for space, such as the one shown in Fig.~\ref{fig:intruder}, can be modelled using game theory. It is assumed that each pedestrian chooses her strategy - her preferred velocity - in order to minimize some cost integrated over the whole future. Each pedestrians will then follow her preferred velocity up to some noise, namely $\dot{\vec X}_t = \vec{v}_0(t) + \sigma \xi(t)$, 
with $\xi$ a $\delta$-correlated normalized white noise.
A possible cost is
\begin{equation}
    \label{eq:cost}
    \begin{aligned}
		c[\vec{v}_0(\cdot)] (\vec{x},t) =  
  \mathbb{E}\bigg[\int_t^T \left[\frac{\mu}{2}(\vec{v}_0(\tau))^2 - g\rho(\vec{X}_\tau,\tau) + U_0(\vec{X}_\tau,\tau)\right] e^{\gamma(t-\tau)}d\tau\\
  +\; e^{\gamma(t-T)}c_T(\vec{X}_T) \bigg]_{|\vec{X}_t = x} \; ,
    \end{aligned}
\end{equation}
where the expectation value is taken on the noise of the Langevin equation.
This cost includes some penalty for high velocities - see the square term $(\vec{v}_0)^2$ 
- and, with $g<0$, for high pedestrian densities $\rho$. Besides, the potential $U_0$, assumed large, prevents pedestrians to occupy the space under obstacles - for example here the cylinder. The last term $c_T$ is a terminal cost which may be useful for example if pedestrians have to reach a given location.  The integral is taken over a long future ($T$ is supposed to be large), but the exponential term allows to bound the anticipation to a typical timescale $1/\gamma$. 

Note that the cost does not depend on the precise pedestrian, but is a function of $\vec{x}$. 
This means that all pedestrians at the same location at the same time will have the same strategy.
The interaction between pedestrians is entirely contained in the density term in the cost, meaning that in this model, pedestrians do not react to the precise location of each other person, but rather to an ensemble averaged density that they foresee, in the spirit of a mean-field assumption. Indeed, if all individuals had to be taken into account, the problem would become overwhelmingly complex.

The OV $\vec{v}^*_0(t)$, that is the velocity that optimized the cost Eq.~\eqref{eq:cost}, can then be computed introducing the {\em value function} $u(\vec{x},t) = \min_{\vec{v}_0(\cdot)} c(\vec{x},t)$ , and  making use of Bellman's optimality principle, which states that if we want to optimize the strategy along its whole history up to time $T$, we need to optimize it between any current time $\tau$ and $T$.
This allows one to construct the history of $u$ backward from the final time $T$, and $u$ is shown to obey the so-called \emph{Hamilton--Jacobi--Bellman} (HJB) equation
\begin{equation}
	\label{eqn:HJB}
	\begin{cases}
		\partial_t u  = -\frac{\sigma^2}{2}\Delta{u} + \frac{1}{2\mu}(\vec{\nabla}u)^2 + \gamma u + g\rho(\vec{x},t) + U_0(\vec{x},t) \;  .\\
		u(\vec{x},t = T) = c_T(\vec{x})
	\end{cases}
\end{equation}
From $u(\vec{x},t )$, the OV is then obtained as $\vec{v}^*_0(t) =  -\vec{\nabla}u/\mu$.
 This equation must be supplemented by the evolution of the density, given that all pedestrians evolve according to the aforementioned \emph{Langevin} equation, with their preferred velocity being the one yielding the minimum cost $u$. The associated equation, forward in time, reads
 \begin{equation}
	\label{eqn:KFP}
	\begin{cases}
		\partial_t \rho = \frac{\sigma^2}{2}\Delta \rho  +\frac{1}{\mu}\nabla\cdot(\rho\nabla u)\\
		\rho(\vec{x},t = 0) = \rho_0(\vec{x})
	\end{cases}
\end{equation}
Solving the two coupled equations \eqref{eqn:HJB}-\eqref{eqn:KFP} can be done iteratively. Besides the equation system can be simplified in some cases, for example when a stationary state is reached.

This framework allows to reproduce the anticipative behaviour of pedestrians, in particular in the case of the incoming cylinder of Fig.~\ref{fig:intruder}. An underlying assumption is that pedestrians can predict the future density, relying on a repeated experience. In case where pedestrians have only a limited knowledge of what is happening - for example because they cannot see some parts of the scene - it can be useful to adjust their anticipation horizon $1/\gamma$~\cite{butano_a_u2024a}.

\subsection{Load balancing in pedestrian evacuation}

\begin{figure}[!ht]
    \bigskip\centering
    \includegraphics[width=0.5\linewidth]{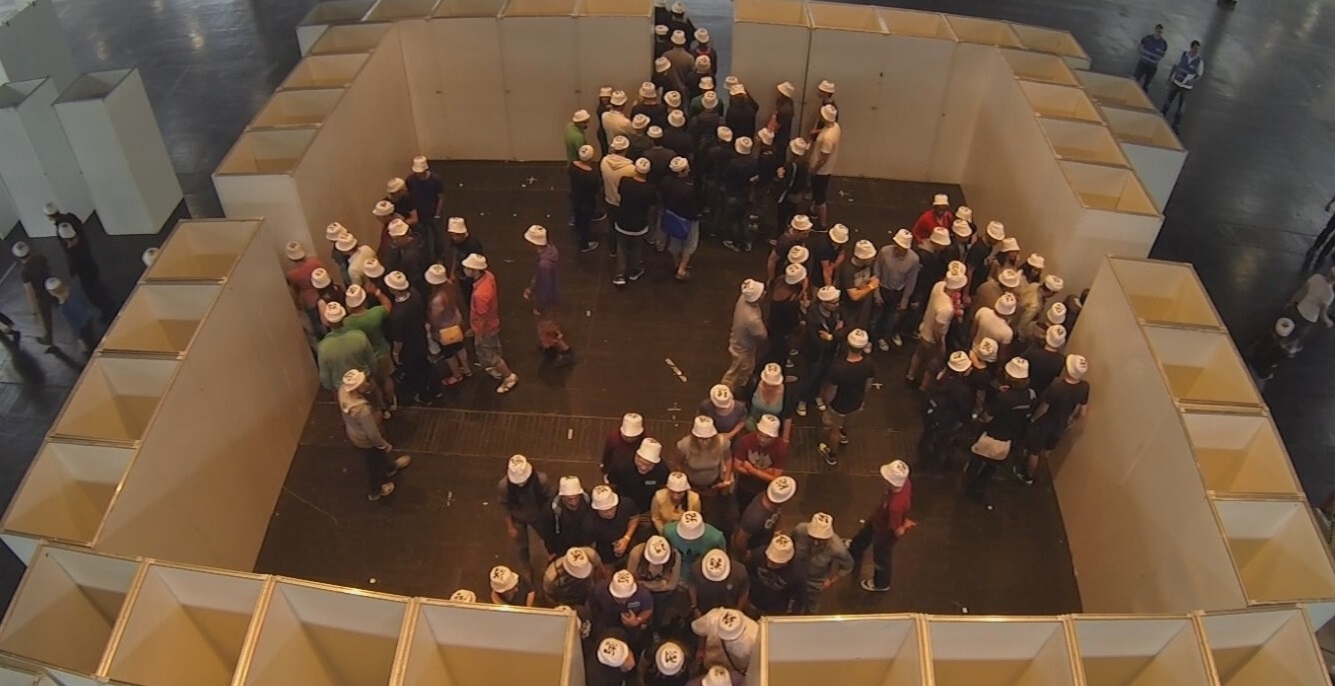}\\[2mm]
    $\qquad$\includegraphics[width=0.35\linewidth]{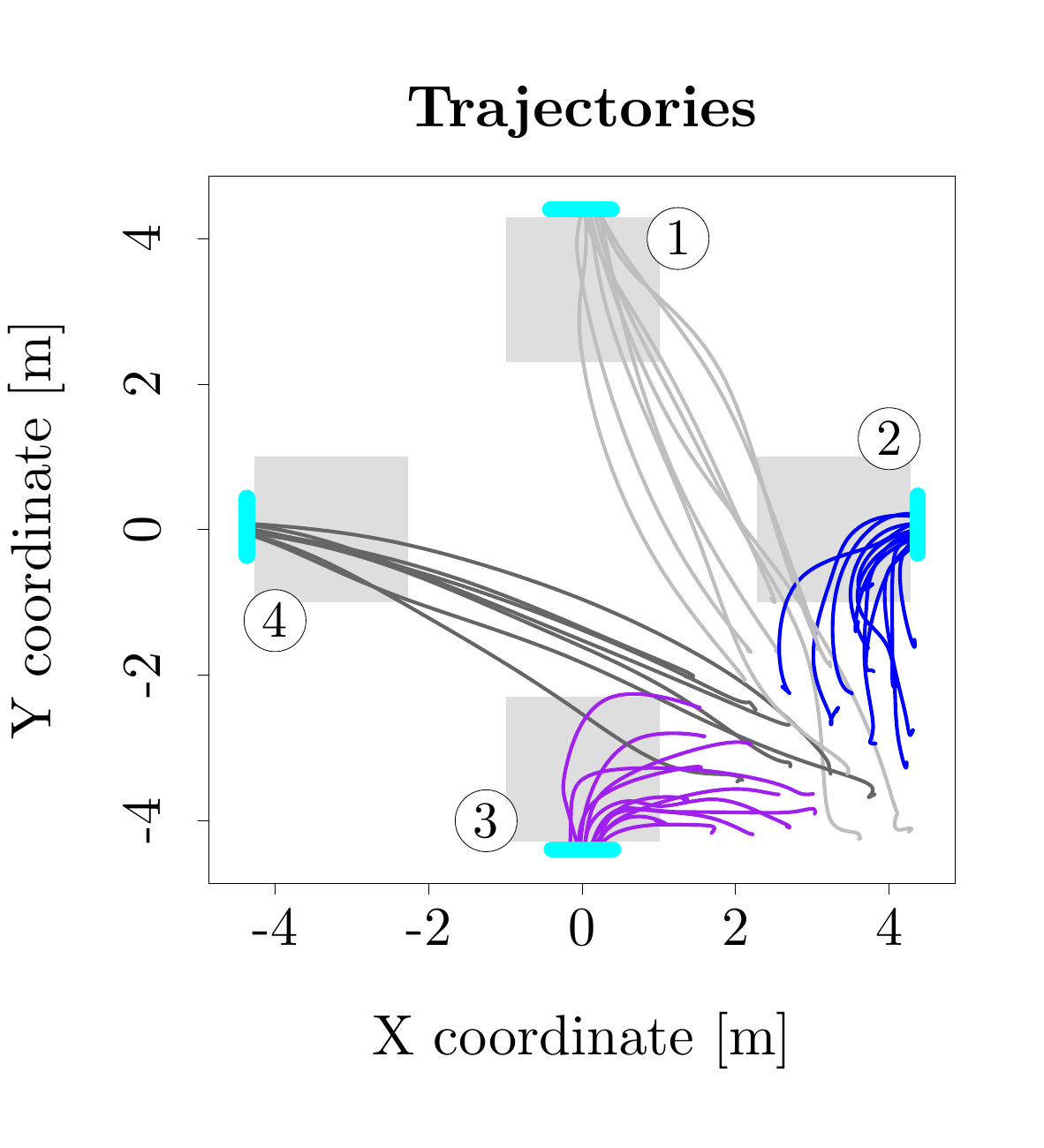}\hspace{1cm}
    \includegraphics[width=0.365\linewidth]{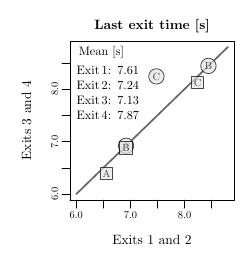}$\qquad$\vspace{-2mm}
    \caption{Pedestrian evacuation experiment with four possible exits \cite{wagoum2017understanding}. Three runs are repeated with pedestrians starting at the bottom right of the geometry (see the trajectories in the lower left panel for the run A). The pedestrian distribution over the four exits shows a load balancing with similar last exit times (see the lower right panel - the circles correspond to exits 2 and 3 while the squares correspond to exits 1 and 4; the letter identifies the run).}
    \label{fig:LoadBalancing}
\end{figure}

Another experiment that reveals collective dynamics and a form of long-term anticipation behaviour is the evacuation of pedestrians under normal conditions in a geometry with multiple exits \cite{wagoum2017understanding}. 
We can observe that the pedestrians spread over the different exits to balance the load and optimise the final evacuation time, whether for rectangular, curved or square geometries, as in Fig.~\ref{fig:LoadBalancing}. 
As the exits are several metres apart, load balancing requires the agent to anticipate for a few seconds to select the exit and move accordingly. 
This typically exceeds the short-term anticipation of the operational model and requires the inclusion of some long-term tactical decision models. 
This involves operational/tactical coupling in modelling the dynamics, which may operate on different time scales.

The tactical choice of an exit is a problem of differential games, because the problem is dynamic and the choice of an exit by a pedestrian depends on the choices of his neighbours. 
In addition, static distance-based shortest path algorithms prove ineffective, as the choice of exit depends on the current level of congestion on the path to the exits.
Real-time pedestrian exit choices can be modelled by centralised optimisation algorithms using, for example, a \emph{Hamilton--Jacobi--Bellman} equation as in a mean field game \cite{hoogendoorn2004pedestrian}, non-cooperative individual behavioural rules \cite{lo2006game,kemloh2012modeling}, or learning processes based on previous experience \cite{crociani2016multidestination}. 
In general, individual estimation of pedestrian travel time can be computationally intensive. 
However, heuristic tactical approaches based on local density levels prove to provide realistic behaviour \cite{kemloh2012modeling,wagoum2017understanding}. 
In any case, the modelling approach requires the tactical exit choice to be dynamic and the operational and tactical modelling scales to be coupled.

\section{Conclusion and challenges}
\label{sec:Challenges}

In summary, we have presented the state-of-the-art in pedestrian and traffic modelling. 
However, many challenges and research directions for modelling the collective dynamics of vehicles and pedestrians still need to be addressed. In conclusion, we now present a few of them. 
These include the stabilisation of road traffic flow, the development of realistic pedestrian behaviour models, the consideration of socio-psychological aspects, or the development of generic order parameters for the characterisation of collective dynamics.

\subsection{Stabilisation of stop-and-go waves}

Stop-and-go dynamics are currently observed in conventional road traffic flow.
In addition, several recent experiments \cite{milanes2014modeling,gunter2020commercially,ciuffo2021requiem,li2021car,makridis2021openacc} show that adaptive cruise control (ACC) driver assistance systems present some stability issues, partly due to large response times of the systems \cite{makridis2019response}. 
In the context of driving automation, there is an important need to develop stable ACC systems that are robust to delays, latency, and stochastic perturbations in the dynamics. 
Indeed, many factors can disrupt the stability of the flow and lead to collective oscillations.

First experiments in the late 2010s show that stable system allows the wave to be dissipated \cite{stern2018dissipation}. 
However, this requires the system to maintain a very large time gap, which limits flow performance. 
Connected platooning systems, known as cooperative adaptive cruise control (CACC) systems, enable efficient automated flows, particularly for truck platooning \cite{milanes2013cooperative,wang2017developing}. 
However, they rely on communication systems and are not autonomous. 
Various compensator modules for lag and delay in the dynamics, typically using feedback control or Smith predictor, as well as filters for noise attenuation are being developed \cite{bekiaris2020pde,xing2020compensation,khound2023unified}. 
However, the topic remains challenging.
Recently, an experiment was conducted in the United States with up to 100 connected and autonomous vehicles to stabilise traffic flows \cite{lee2024traffic}, including reinforcement learning techniques to damp oscillations 
\cite{jang2024reinforcement}\footnote{See also \href{https://circles-consortium.github.io/}{\texttt{https://circles-consortium.github.io/}}}.

\subsection{Developing realistic microscopic pedestrian models}
The practical demand for better pedestrian dynamics models that would superside existing software
suggests that one should strive for as much realism in these models as possible, hence greater sophistication.
A seemingly natural conclusion would be to try and refine the description of agents, notably by better accounting for the psychological processes at play, for the motivation experienced by people and their emotions. Indeed, using the pedestrian flow at a bottleneck as an illustration, it has been shown that the crowd's level of motivation impacts how densely people pack in front of the bottleneck \cite{adrian2020crowds}, which in turn largely controls the flow rate at not-too-high pressure \cite{nicolas2017pedestrian}. 
In this vein, Thalmann and Musse had thus prompted the idea that the interplay between perception and emotion determines an agent's behaviour, and should thus be accounted for \cite{thalmann2012crowd}.

Nonetheless, Corbetta et al.'s ability to reproduce the statistics of U-turns on a staircase platform with a simple scalar potential energy description \cite{corbetta2017fluctuations}, irrespective of the great variety of decisions that led to such U-turns in practice, demonstrates that the dynamical effect of complex psychological factors can be captured at a high level without delving into the substance of these factors. 

In short, the required degree of realism obviously depends on the practical purpose, and in particular whether 
the consequences of the motivation experienced by people (in terms of desired speed, direction, etc.) can be taken as an input to the model, or need to be explained in greater depth.
That being said, for any practical application, we believe that state-of-the-practice models can hardly 
ignore two long-overlooked factors: 
(i) the presence of social groups (of friends, co-workers, 
family, etc.), which implies a unique 
relative arrangement of people in space (taking up more lateral 
space on a corridor than would be predicted for single individuals), (ii) the non-uniformity of space as 
perceived by pedestrians in real settings, which sharply contrasts with 
the uniform space idealised in controlled experiments designed to single out intrinsic features of crowd dynamics.


\subsection{Including socio-psychological concepts}

In many situations, a natural scientific perspective on the dynamics of pedestrians is not sufficient, but must be supplemented by sociological or psychological concepts, such as, for instance, group behaviour, social identity \cite{seitz2017parsimony,wijermans2022towards} or social norms \cite{siebenbehavioral}. This is illustrated below with two examples. 

In some situations, one can observe different behaviour of individuals which can change dynamically. In \cite{sieben2017collective,adrian2020crowds} this can be observed in the time distance plots, Fig. 6 in \cite{adrian2020crowds} where the paths show that some people push their way to the front while others fall back. A more detailed analysis is provided in \cite{usten2022pushing} where psychologist developed an individual annotation of the behaviour of people in a crowd. These behavioural data allow to study the space-time dynamics of pushing behaviour like the relation of distance to the entrance and the probability of pushing, see \cite{uesten2023exploring}. The authors showed, for example, that the probability to observe pushing behaviour increases when people approach their goal, in this experiment the entrance to a concert. 

To reproduce such a space-time dynamic of behaviour, state-of-the-art models need layers that model behavioural changes. Furthermore, a critical look at classic movement models in this respect makes it clear that formulations of the interaction of the particles are mostly designed to model a mixture of behaviours, e. g., a force model is able to describe the transfer of forces when people push and is also used to model avoidance manoeuvres when crossing other people. Here, the models quickly reach their limits when it comes to realistically representing the diversity of behaviours in a single simulation.

A further example is the behavioural diversity and dynamic as described in \cite{siebenbehavioral}. The authors study a entrance situation with different motivation, where queuing but also pushing could be observed. An ethogram is crated to document the diversity of behaviour and it is shown how pedestrians change their behaviour with time, see Fig. 7. The formation of a queue and its meaning as a social rule leads to a collective phenomenon for which many people work together towards a common goal. Pairwise interaction schemes such as the superposition mechanisms in force-based models corresponding to weighted averages are too simplistic and cannot consider the collective aim to form a queue. This also applies to common velocity models where the interaction with the neighbour is helpful to model the avoidance of collision but are not usable to consider the formation of an order by queuing. Furthermore, current models are unable to take into account the triggers for behavioural changes documented in \cite[Fig.~7]{siebenbehavioral}.

To describe collective interaction concepts from psychology and sociology should be integrated in modelling of the movement of agents. These could be sociological concepts such as norms and scripts \cite{siebenbehavioral} but also psychological concepts such as perception \cite{wirth2023neighborhood}, cognition or the influence of leadership \cite{lombardi2020nonverbal}.

Pairwise interaction schemes, such as the superposition mechanisms in force-based models or single interaction with the neighbour with the shortest time-to-collision in certain velocity-based models, are too simplistic to account for complex behavioural changes. 
In fact, such mechanisms rely on intricate collective interaction schemes based on sociological and psychological cognitive aspects, and complex threshold mechanisms \cite{gregorj2023social,zhang2023review}. 
Other aspects rely on safety components, especially for vulnerable users such as pedestrians and cyclists in mixed urban traffic \cite{ni2016evaluation}. 
The modelling of safety behaviours is based on pedestrian risk assessment that can affect the collective dynamics of a crowd and mixed traffic flow \cite{shen2021pedestrian,zhang2021pedestrian}.
In general, considerable efforts are still needed to identify the sociological, psychological and safety aspects involved and to model mechanisms capable of taking them into account \cite{beermann2023connection,templeton2020modeling,adrian2020crowds,amini2022development}.



\subsection{Developing generic order parameters}
Another important challenge is to quantify and characterise the collective dynamics of vehicles and pedestrians using aggregated order parameters. 
In general, specific order parameters are defined for each of the collective dynamics. 
For example, stop-and-go waves can be characterised using variability indicators of vehicle speeds or gaps, while lane formation can be determined by counting the number of pedestrians in front with the same direction \cite{nowak2012quantitative}. 
However, the formulation of generic order parameters remains challenging.

Lately, a modelling paradigm for particle, vehicle and pedestrian dynamics based on port-Hamiltonian systems has been developed \cite{knorn2014passivity,matei2019inferring,tordeux2023multi,ehrhardt2024collective,rudiger2024stability}. 
Port-Hamiltonian systems (PHS) are a recent but already well-established modelling approach for nonlinear physical systems \cite{van2006port}. 
They decompose the dynamics in between distance-based repulsion corresponding to the Hamiltonian part with energy conservation, and some relaxation to a tactical desired velocity corresponding to the dissipative part of the system and to the input ports. 
Here, the Hamiltonian can be used as a generic order parameter characterising the collective dynamics \cite{tordeux2023multi}. 
In addition, the Hamiltonian can be used as a Lyapunov function to address the system stability thanks to the dissipativity inequality without the need to linearise the system. 
The Hamiltonian in PHS is a promising modelling approach to quantify collective dynamics in a generic way. 
However, the subject remains challenging and is still under development.


\bibliographystyle{abbrv}
\bibliography{refs}

\end{document}